\documentclass[letterpaper,floatfix,showpacs,aps,prl,twocolumn,10pt,superscriptaddress,showpacs]{revtex4}

\usepackage{latexsym,amsmath,amssymb,amsfonts,mathbbol,graphicx,color}
\usepackage{dcolumn}% Align table columns on decimal point
\usepackage{bm}% bold math

\newcommand{\ket}[1]{|{#1}\rangle}
\newcommand{\bra}[1]{\langle {#1}|}
\begin{document}

\title{Quantum pseudo-randomness from cluster-state quantum computation}

\author{Winton G. Brown}
\affiliation{Department of Physics and Astronomy, Dartmouth
College, Hanover, NH 03755, USA}

\author{Yaakov S. Weinstein}
\affiliation{Quantum Information Science Group, {\sc Mitre},
260 Industrial Way West, Eatontown, NJ 07224, USA}

\author{Lorenza Viola}
\affiliation{Department of Physics and Astronomy, Dartmouth
College, Hanover, NH 03755, USA}

\date{\today}

\begin{abstract}
We show how to efficiently generate pseudo-random states suitable for quantum information processing via cluster-state quantum computation.  By reformulating pseudo-random algorithms in the cluster-state picture, we identify a strategy for optimizing pseudo-random circuits by properly choosing single-qubit rotations.  A Markov chain analysis provides the tool for analyzing convergence rates to the Haar measure and finding the optimal single-qubit gate distribution.  Our results may be viewed as an alternative construction of approximate unitary $2$-designs.
\end{abstract}

%24.10.Cn
\maketitle

Methods for characterizing and efficiently generating random quantum states and unitary operators have broad conceptual and practical significance across quantum physics.  From a fundamental standpoint, a main motivation stems from the challenge of modeling complex quantum behavior, including quantum chaos \cite{Haake} and typical entanglement in many-body systems
\cite{Page,WH2,Patrick,Dahlsten,Znidaric,MSB,Brown}.  Within quantum information science, states and unitaries sampled from the appropriate uniform (Haar) distribution provide the enabling resource in a growing number of algorithms and protocols.  Remarkably, random pure states saturate the classical communication capacity of a noisy quantum channel \cite{Seth2}, and allow superdense coding of arbitrary quantum states \cite{HHL}. Random unitaries find applications in tasks ranging from approximate encryption and remote state preparation \cite{HLSW}
%\cite{Bennett},
to unbiased noise estimation \cite{Science,Levi} and selective process
tomography \cite{Paz}.

However, implementing exact randomization on a quantum computer is
inefficient, as the number of required elementary gates grows exponentially with the number of qubits.  Still, it has been shown \cite{Science,Dahlsten,Convergence} that one can generate {\em pseudo-random} (PR) quantum states and unitary operators which satisfy certain practical tests of randomness using only a polynomial number of gates.  In particular, a framework for quantifying to what extent pseudo-randomness may simulate the Haar distribution for an intended randomization task is offered by the notion of a {\em $t$-design} \cite{EmersonDesign}.  In its essence, a state (unitary) $t$-design is a probability distribution over pure states (unitaries) whose statistical moments up to order $t$ equal those from the Haar distribution.  While efficient exact unitary $2$-designs are known \cite{EmersonDesign},  constructions of {\em approximate} $2$-designs as well as alternative schemes tackling {\em higher-order} moments are actively investigated \cite{Dahlsten,Znidaric,MSB}.  So far, existing studies have focused only on the circuit model of quantum computation (QC).

In this work, we construct an efficient algorithm for PR state generation in the {\em cluster-state} paradigm of QC \cite{BR1}.  This is crucial from an implementation perspective, in that many of the above-mentioned applications of random states originate in quantum communication protocols, for which photonic entanglement and cluster-state QC provide a leading approach \cite{N}. Furthermore, we find that reformulating PR algorithms in a cluster-state picture suggests a path to optimize existing circuit constructions.  In particular, by analyzing PR circuits in terms of classical Markov chains \cite{Dahlsten}, we identify an {\em optimal} single-qubit gate distribution -- complementing existing results on optimal two-qubit gates \cite{Znidaric}.  Quantitative convergence bounds are obtained by invoking standard tools from spectral mixing analysis.

\textit{Pseudo-Random Quantum Circuits.$-$}
A PR circuit on $n$ qubits attempts to generate states and unitaries whose statistical properties mimic those of Haar-distributed counterparts by
repeated applications of random one- and two-qubit gates.  In the PR algorithm of \cite{Science}, single-qubit gates drawn uniformly from the Haar measure on SU$(2)$ are performed in parallel on each qubit, followed by a controlled phase (CZ) gate on all nearest-neighbor pairs.  Iterating {\em any} quantum circuit constructed from a universal set of random gates eventually converges to the Haar measure as the circuit depth increases \cite{Convergence}.  However,
%because a generic element in U($2^n$) cannot be constructed efficiently,
the rate of convergence for some test functions which probe arbitrarily high-order moments of the PR-distribution may scale exponentially with $n$. Thus, PR algorithms can adequately reproduce the Haar measure with polynomial effort only for a restricted class of test functions.  This may suffice for practical applications as long as the quantities of interest are known to involve only low-order moments, as in $t$-design approaches \cite{Dahlsten}.

In our analysis, we select two illustrative test functions, and track them as a function of iteration.  The first is the distribution of the squared moduli of
state-vector components, $P(\eta)$, in the computational basis. For random pure states, this is given by the Porter-Thomas distribution, $P_{PT}(y) = \exp(-y)$,
where $y = N\eta$, $N=2^n$ \cite{Zyc}. Accordingly, we examine the
$l_2$-distance between $P_{PT}(y)$ and the PR state distribution.  The
second test function we employ is the average subsystem entanglement,
expressed in terms of the Meyer-Wallach entanglement measure \cite{MW,WH3},
$Q = 1-\frac{1}{n}\sum_{i=1}^n \sum_{a = 0,x,y,z} \bra{\psi}\sigma_a^i\ket{\psi}^2,$ which quantifies the average purity of single-qubit reduced density matrices.
For a random pure state, the expected value of $Q$ is $\mathbb{E}(Q)\equiv  Q_R = (2^n-2)/(2^n+1)$. Within the approach of generalized entanglement \cite{Barnum}, $Q$ is a representative of a class of quadratic measures of state delocalization and generalized purities \cite{Brown}, for which a similar analysis may be developed.

\textit{Pseudo-Random Cluster-State Computation.$-$} Cluster states
are highly entangled states which serve as the basic resource for
measurement-based QC \cite{BR1}. They may be generated by applying CZ gates between qubits initially in the $\ket{+}$ state.  Computation is executed by measuring qubits along desired axes in the $x$-$y$ plane.  The choice of measurement axis determines the operation that is implemented, and may depend on the outcome of a previous measurement.  A 2D qubit lattice with CZ gates applied between all nearest neighbors suffices for universal QC. Measurements are performed by column from left to right, until a last column is left unmeasured -- determining the statistics over the computation outcomes.

\vspace*{-3mm}
\begin{figure}[h]
\begin{center}
\begin{picture}(250,60)(-25,40)
\setlength{\unitlength}{.8pt}
\multiput(20,65)(32,0){7}{\circle{16}}
\multiput(20,105)(32,0){7}{\circle{16}}
\multiput(28,65)(32,0){6}{\line(1,0){16}}
\multiput(28,105)(32,0){6}{\line(1,0){16}}
\put(17,103){$\alpha_1$}\put(49,103){$\beta_1$}\put(81,103){$\gamma_1$}
\put(17,63){$\alpha_2$}\put(49,63){$\beta_2$}\put(81,63){$\gamma_2$}
\put(113,103){$\delta_1$}\put(145,103){$\epsilon_1$}\put(177,103){$\zeta_1$}
\put(113,63){$\delta_2$}\put(145,63){$\epsilon_2$}\put(177,63){$\zeta_2$}
\put(116,75){\line(0,1){20}}
\put(212,75){\line(0,1){20}}
\multiput(19,44)(32,0){7}{\vdots}
\multiput(52,76)(0,5){4}{\line(0,1){3}}
\multiput(84,76)(0,5){4}{\line(0,1){3}}
\multiput(148,76)(0,5){4}{\line(0,1){3}}
\multiput(180,76)(0,5){4}{\line(0,1){3}}
\end{picture}
\end{center}
\vspace*{-3mm}
\caption{Schematics of cluster-state PR architectures.
Pairs of qubits subjected to CZ gates are connected by solid lines
and each qubit is identified by the angle in the $x$-$y$ plane that defines its measurement basis. Dashed lines represent additional CZ gates for the enhanced version of the algorithm.}
\label{cluster}
\end{figure}
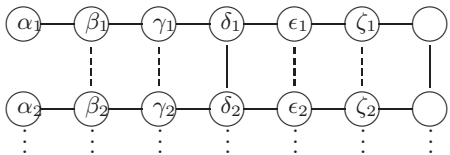

A cluster state architecture that implements the equivalent of $2$ iterations of the PR algorithm of \cite{Science} is depicted in Fig.~\ref{cluster}. Using Euler-angle representation, measurement of $3$ qubits in a row simulates a single-qubit gate
$HZ(\alpha_i+\pi m_{\alpha_i})X(\beta_i + \pi m_{\beta_i}) Z(\gamma_i
+ \pi m_{\gamma_i})$, where $H$ is the Hadamard gate, $Z(\alpha)$
($X(\alpha$)) is a $z$- ($x$-) rotation by an angle $\alpha$,
$(\alpha_i, \beta_i, \gamma_i)$ are the angles along which each qubit is
measured in the $x$-$y$ plane, and $m_i=0,1$ labels additional
measurement-dependent $\pi$ rotations. Arbitrary single-qubit
gates are effected by properly choosing the Euler angles.  Because the Haar measure on SU$(2)$ is invariant under the extra $\pi$ rotations, the latter may be ignored.  CZ gates performed between rows of the cluster state (vertical lines) serve as CZ gates acting between qubits in the circuit-based algorithm. In general, to simulate $\ell$ iterations of an $n$-qubit PR circuit, a lattice of $n\times 3\ell+1$ qubits is needed (where the extra
$1$ comes from the final, unmeasured column). The first column contains the initial state $|\psi_0\rangle$ on which the algorithm operates.  While the latter can be arbitrary in principle, we always set $|\psi_0\rangle=\ket{0\ldots 0}$ for PR state generation.

\begin{figure}[t]
\centerline{
\includegraphics[width=2.9in]{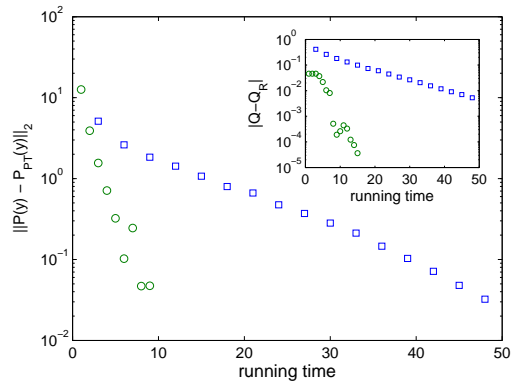}}
\caption{(Color online) Distance of the (normalized) distribution of
squared moduli state-vector components from $P_{PT}(y)$
for random states as a function of run time.
$\square$: Standard ($6$ rows and connections every third column), vs
$\bigcirc$: Enhanced (connections at every column) PR patterns.
The run time equals the number of columns in the cluster state.
Inset: Difference of global entanglement, $Q$, from the expected
random-state value, $Q_R$, vs run time. For both test functions, the enhanced version of the algorithm converges to the Haar average with a rate about $6$ times faster than the standard rate. }
\label{components}
\end{figure}

Given the measurement pattern of Fig.~\ref{cluster}, a natural question arises: Can the convergence rate of the cluster-state PR algorithm be enhanced by filling in {\em additional} vertical lines (that is, by effecting additional CZ gates represented by dashed lines)? In this case, QC proceeds as before but measurement angles will be chosen randomly in the $x$-$y$ plane.  Fig.~\ref{components} illustrates the resulting improvement by comparing the decay rate of both $P(y)$ and $Q$: For 6 row cluster states, both test functions converge approximately $6$ times faster for the completely filled cluster state. Thus, the enhanced version of the PR cluster-state algorithm uses a factor of 6 fewer qubits and horizontal connections, and half the number of vertical connections to achieve a comparable distance from random-state behavior.

\textit{Enhanced Pseudo-Random Circuit Design.$-$}  Because the cluster model is computationally equivalent to the circuit model \cite{BR1}, the improvement observed for the enhanced cluster-state PR algorithm should have a circuit model analog. The single-qubit rotation equivalent to measuring a cluster qubit in a random basis in the $x$-$y$ plane is an $HZ(\alpha)$ gate.  Thus, once translated into the circuit model, the completely filled measurement pattern identifies a {\em restricted} family of random single-qubit gates which map the $z$-axis to the transverse plane.

\begin{figure}[t]
\centerline{ \includegraphics[width=2.9in]{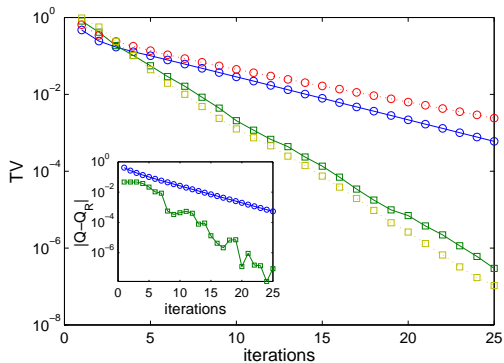}}
\caption{(Color online) Total variation distance,
$TV(\ell) = {1}/{2}\sum_\nu |c_\nu^2 (\ell) - c_\nu^2(\infty)|$,
as a function of iteration for a PR circuit with random single-qubit
gates (circles) and $HZ(\alpha)$ gates (squares).  Data is shown for
$n=6$ (solid lines) and $n=12$ (dashed lines).
Inset: Difference of global Meyer-Wallach entanglement from the
expected random-state value as a function of iteration for a PR
map on $n=6$ qubits.}
\label{TV}
\end{figure}

Why should such a restriction improve the convergence rate? The answer
has to do with the relationship between the one and two-qubit gates comprising the algorithm. Single-qubit rotations that preserve the $z$-axis \textit{commute} with CZ gates, thus they do not aid in scrambling the quantum state. The $HZ(\alpha)$ gates are non-invariant with respect to the $z$-axis. The inset in Fig. \ref{TV} contrasts the convergence to typical entanglement in PR circuits with fully random and restricted single-qubit gates. In the circuit model, the decay rate of the enhanced algorithm is only about $2$ times faster, since a time step counts as a complete iteration. Still,
depending on implementation, it may be easier to perform an $HZ(\alpha)$ gate than an arbitrary single-qubit gate.  This raises the following general question: Given a {\em fixed} two-qubit gate, what is the {\em optimal}
single-qubit gate distribution to employ?  The key insight is to relate convergence properties of a PR circuit to those of an appropriate random walk
\cite{Dahlsten,Znidaric}.

\textit{Markov Chain Analysis.$-$}
As established in \cite{Dahlsten}, the idea is to map the evolution of the second moments of the final state under PR-unitaries to a classical Markov chain.  Let a $n$-qubit density operator be expressed in the Pauli basis: $\rho=
%\ket{\psi}\bra{\psi} =
\sum_\nu c_\nu P_\nu$, where $P_\nu$
% =\sigma_1^{\nu_1} \ldots \sigma_n^{\nu_n}$
is a tensor-product string of single-qubit identity and Pauli operators,
specified by the collective index $\nu \in {\mathcal I}=\{0,x,y,z \}^n$.
The coefficients $\{c_\nu^2 \}$ form a probability distribution over
${\mathcal I}$. Let PR$(\ell)$ be the family of PR circuits of depth
$\ell$. The ensemble-averaged coefficients
$\{{\mathbb E}_{\text{PR}(\ell)}(c_{\nu,\ell}^2 )\}$ also form a probability distribution over ${\mathcal I}$.
Under conditions described below, the rules for updating these coefficients
follow a discrete-time Markov chain on ${\mathcal I}$.  That is, if the chain is initially distributed according to $\{c_{\nu,0}^2 \}$, the distribution of the evolved state satisfies
$${\mathbb E}_{\text{PR}(\ell+1)}(c_{\nu,\ell+1}^2)=
\sum_{\nu \in {\cal I}} M_{\mu \nu}
{\mathbb E}_{\text{PR}(\ell)}(c_{\nu,\ell}^2)=
\sum_{\nu \in {\cal I}} M_{\mu \nu}^\ell c_{\nu,0}^2, $$
where $M=\{M_{\mu \nu}\}$ is the transition matrix of chain.

Clearly, it suffices to construct $M$ for a single iteration. Under a single-qubit gate in SU(2), each non-trivial Pauli operator transforms as $\sigma_a\mapsto R(\sigma_a)=\sum_{b} x_{ab}\sigma_b,$ $a,b \in \{x,y,z\}$, where $R=\{x_{ab}\}\in$ SO(3). The corresponding
$4\times 4$ Markov matrix $\overline{R}$ is obtained by averaging the squared coefficients over the distribution of local gates.  $\overline{R}$ can only be constructed when the ensemble averages of each cross term, ${\mathbb E}(x_{ab}x_{ac})$ with $b\neq c$, vanishes, leaving $\overline{R}_{ab} ={\mathbb E}(x_{ab}^2)$. Because the single-qubit gates on different qubits are selected independently, the transformation resulting from the overall local part of the PR map is the $n$-fold tensor product of single-qubit transformations,
$L = \overline{R}^{\otimes n}$. Since each CZ gate preserves the Pauli group and acts, up to phases, as a permutation on the columns of $L$, the full transformation $M$ is a Markov matrix if $\overline{R}$ is.

In order to identify the optimal single-qubit gate distribution, we take advantage of the fact that CZ gates do not distinguish between the $x$, $y$ axes, and restrict to distributions which initially randomize states in the $x$-$y$ plane. This allows the construction of a {\em reduced Markov chain}, whose transition matrix $M'$ has exactly the same (non-zero) eigenvalues
as $M$.  Let $P$ be a Pauli operator containing at least one $X_i$ or $Y_i$, and let $P'$
be any operator obtained from $P$ by permuting $X_i$ with $Y_i$. Since $M$ randomizes $X_i$
and $Y_i$, $M(P-P') = 0$. This defines the kernel of $M$, which may be removed by defining
new variables $\Xi_i^\pm = X_i \pm Y_i$.
Chain states including {$\Xi_i^-$} may be discarded, whereas transitions within ${\cal I}'= \{0,z,\xi\}^n$ are described by $M'$. Let $c \in [0,1]$ parametrize the extent to which the $z$-axis is left invariant.  The single-qubit gate contribution to $M'$ is then fully described by:
%\cite{remark2}:
\begin{eqnarray*}
R(c)= \left(
  \begin{array}{cccc}
    1 & 0 & 0  \\
    0 & c & \frac{1-c}{2} \\
    0 & 1-c & \frac{1+c}{2} \\
  \end{array}
\right).
\end{eqnarray*}

For the PR circuits examined here, the Markov chain obtained by
removing the identity (that is, by restricting to ${\cal I}'-\{0\}^n$)
is ergodic, with a stationary distribution corresponding to the uniform
distribution on ${\cal I}$.  The convergence is asymptotically exponential, with a
rate $ \Gamma(c)$ determined by the gap between the largest and second
largest eigenvalues, $\Delta(c) = \lambda_1 - \lambda_2$, via
$ \Gamma(c) = - \ln(1-\Delta(c))$. Thus, the larger the gap the
faster the convergence. Here, we are interested in the gap between
$1$ and the next largest eigenvalue whose eigenvector has a
non-zero component along $|\psi_0\rangle\langle\psi_0|$.

\begin{figure}[t]
\centerline{
\includegraphics[width=2.8in]{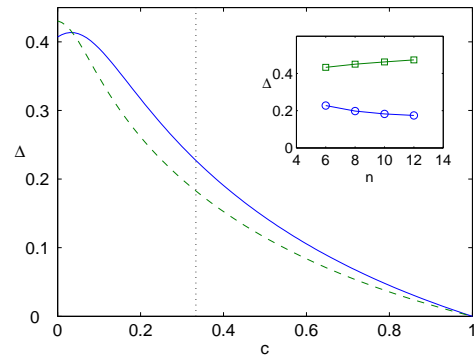}}
\caption{(Color online) Gap, $\Delta(c)$, between $ \lambda_1 =1$
and the largest non-unit eigenvalue of $M'$ vs $c$, for $n=6$
(solid line) and $n=10$ (dashed line). The gap at $c = 0$ ($HZ(\alpha)$
gates) is significantly larger than the gap at $c = 1/3$ (arbitrary
random single-qubit gates), identifying the optimal gate set.
Inset: Gap for $c={1}/{3}$ (circles) and $c=0$ (squares) vs $n$. }
\label{gap}
\end{figure}

As seen in Fig.~\ref{gap}, the maximum gap for $n=6$ qubits, equal to
$0.4135$, occurs at $c \approx 0.03$. The gap for $HZ(\alpha)$ gates,
$\Delta(0)\approx 0.4071$, is significantly larger than the gap for
random rotations, $\Delta (1/3) \simeq 0.2292$, yielding
$\Gamma(0)/\Gamma(1/3) \approx 2.008$. The agreement of this ratio with the data is clear from Fig.~\ref{TV}, which shows the decay rate of the total variation distance (TV) (that is, the $l_1$-distance) between the distribution undergoing the Markov process and the asymptotic distribution.
For $n=10$, the maximum gap is attained at $c=0$. Remarkably, as $n$ increases, $\Delta$ {\em decreases} for unrestricted local gates, but {\em increases} for $HZ(\alpha)$ gates, see inset of Fig. \ref{gap}.  Thus, the larger $n$, the faster the Markov chain converges.  While determining the asymptotic behavior of the gap is beyond our current scopes, this feature may prove advantageous for small-scale PR implementations.

\textit{Cut-Off Behavior.$-$} While an ergodic Markov chain is guaranteed to exponentially approach stationarity at long times, a practical question is to determine how soon the exponential regime is entered.  In \cite{Dahlsten}, it was found that the \textit{distribution} of entanglement
over an ensemble of PR states remains far from the asymptotic distribution until a time $\tau$ is reached, numerical data suggesting that such a {\em cut-off effect} \cite{Diaconis} becomes sharper as $n$ increases.  In our case, Fig.~\ref{TV} shows no cut-off behavior in the TV for either $HZ(\alpha)$ gates or random single-qubit gates. Increasing $n$ modifies the slope of the TV decay, but does not engender a cut-off phenomenon.  Conclusions on the entanglement distribution are more delicate, as the Markov chain only describes convergence of the lowest entanglement moment. As seen in the inset, exponential convergence occurs immediately for
$Q$ in a circuit using random single-qubit gates. Under $HZ(\alpha)$ gates, however, $Q$ is maximum for the first $n/2$ iterations before exponential decay sets in \cite{remark3}. Thus, every realization of such a PR algorithm
%(starting from $|\psi_0\rangle$)
is maximally entangled for the first $n/2$ iterations, implying that the resulting entanglement distribution is singular until $\tau = n/2$.
%thus very far from that of random states.
This suggests a cut-off effect in the entanglement distribution of enhanced PR circuits, although not in the Markov chain describing the evolution of quadratic test functions.

\textit{Discussion.$-$}
Our results on the emergence of typical entanglement may naturally be
viewed as probing to what extent PR circuits approximate a unitary
$2$-design. This follows from the fact that for every continuous local gate
distribution, there is a discrete distribution over gates belonging to the
single-qubit Clifford group which has the same corresponding Markov matrix.
Thus our procedure can then be viewed as implementing a biased sampling from
the $n$-qubit Clifford group, which is an exact unitary $2$-design
\cite{EmersonDesign}.  Beyond their use as approximate
$2$-designs, however, the PR algorithms studied here are also
approximate $t$-designs for some $t>2$, since they converge to the
Haar measure.  While the fact that PR circuits yield approximate
$2$-designs has been formally proved very recently in \cite{Harrow08},
establishing mathematical and physical connections between PR algorithms
and approximate $t$-designs of higher order is an important next step
toward harnessing quantum pseudo-randomness.

We thank D.G. Cory, C.S. Hellberg, and C.C. Lopez for
feedback.  WGB acknowledges partial support from C. and
W.  Burke through their Special Projects Fund in QIS.  YSW
acknowledges support from the MITRE Technology Program under MTP
grant \#07MSR205.


\vspace*{-2mm}

\begin{thebibliography}{99}

\vspace*{-1mm}

\bibitem{Haake}
F. Haake, {\it Quantum Signatures of Chaos} (Springer, New York, 2001).

\bibitem{Page}
D.N. Page, Phys. Rev. Lett. {\bf 71}, 1291 (1993).

\bibitem{WH2}
Y.S. Weinstein and C.S. Hellberg, Phys. Rev. Lett. {\bf 95}, 030501 (2005).

\bibitem{Patrick}
%Generic entanglement
P. Hayden, D.W. Leung, and A. Winter, Commun. Math. Phys. {\bf 265}, 95 (2006).

\bibitem{Dahlsten}
R. Oliveira, O. Dahlsten, and M. Plenio, Phys. Rev. Lett. {\bf 98},
130502 (2007); O. Dahlsten, R. Oliveira, and M. Plenio, J. Phys. A
{\bf 40}, 8081 (2007).

\bibitem{Znidaric}
M. Znidaric, Phys. Rev. A {\bf 76}, 012318 (2007).

\bibitem{MSB}
Y. Most, Y. Shimoni, and O. Biham, Phys. Rev. A {\bf 76}, 022328 (2007).

\bibitem{Brown}
L. Viola and W. G. Brown, J. Phys. A {\bf 40}, 8109 (2007); W. G. Brown
{\em et al.},
%D. Starling, L.F. Santos, and L. Viola,
Phys. Rev. E {\bf 77}, 021106 (2008).

\bibitem{Seth2}
S. Lloyd, Phys. Rev. A {\bf 55}, 1613 (1997).

\bibitem{HHL}
%Superdense coding
A. Harrow, P. Hayden, and D.W. Leung, Phys. Rev. Lett. {\bf 92},
187901 (2004).

\bibitem{HLSW}
%Approx encryption
P. Hayden {\em et al.},
%, D.W. Leung, P.W. Shor, and A. Winter,
Commun. Math. Phys.
{\bf 250}, 371 (2004);
A. Ambainis and A. Smith, {\it Lect. Notes Comp. Science}
{\bf 3122}, 249 (2004).
%\bibitem{Bennett}
%Remote state prep
C.H. Bennett {\em et al.},
%, P. Hayden, D.W. Leung, P.W. Shor, and A. Winter,
IEEE Trans. Inf. Theory {\bf 51}, 56 (2005).

\bibitem{Science}
J. Emerson {\em et al.},
%, Y.S. Weinstein, M. Saraceno, S. Lloyd, and D.G. Cory,
Science {\bf 302}, 2098 (2003).

\bibitem{Levi}
B. Levi {\em et al.},
%, C.C. Lopez, J. Emerson, and D.G. Cory,
Phys. Rev. A {\bf 75}, 022314 (2007);
J. Emerson {\it et al.}, Science {\bf 317}, 1893 (2007).

\bibitem{Paz}
A. Bendersky, F. Pastawski, and J.P. Paz, {\tt quant-ph/ 0801.0758}.

\bibitem{Convergence}
J. Emerson, E. Livine, and S. Lloyd, Phys. Rev. A {\bf 72},
060302 (2005).

\bibitem{EmersonDesign}
C. Dankert {\em et al.},
%, R. Cleve, J. Emerson, and E. Livine,
{\tt quant-ph/0606161};
A. Ambainis and J. Emerson, {\tt quant-ph/0701126}.

\bibitem{BR1}
R. Raussendorf and H.J. Briegel, Phys. Rev. Lett. {\bf 86}, 5188 (2001);
R. Raussendorf, D.E. Browne, and H.J. Briegel, Phys. Rev. A {\bf 68},
022312 (2003).

\bibitem{N}
M.A. Nielsen, Phys. Rev. Lett. {\bf 93}, 040503 (2004); P. Walther
{\em et al.}, Nature {\bf 434}, 169 (2005); R. Prevedel {\em et al.},
J. Opt. Soc. Am. B {\bf 24}, 241 (2007).


\bibitem{Zyc}
Following standard practice, $y$ is rescaled to have unit mean, and
the limit $N\rightarrow\infty$ is taken. See e.g. F. Haake and K.
Zyczkowski, Phys. Rev. A {\bf 42}, R1013 (1990).

\bibitem{MW}
D.A. Meyer and N.R. Wallach, J. Math. Phys. {\bf 43}, 4273 (2002).
%; G.K. Brennen, Quant. Inf. Comp., {\bf 3}, 619, (2003).

\bibitem{WH3}
These two properties are related as outlined in
Y.S. Weinstein and C.S. Hellberg, Phys. Rev. A {\bf 72}, 022331 (2005).

\bibitem{Barnum}
H. Barnum {\em et al.},
%, E. Knill, G. Ortiz, R. Somma, and L. Viola,
Phys. Rev. Lett. {\bf 92}, 1070902 (2004).

\iffalse
\bibitem{remark1} Such gates may be written as $Z(\beta)HZ(\alpha)$.  However, upon repeated iterations of random $Z(\beta)HZ(\alpha)$ gates, one of the two $z$-rotations is redundant, thus we may substitute gates of the form $HZ(\alpha)$.
\fi

%\bibitem{remark2} Not every single-qubit gate distribution admits a
%Markov chain description.  A sufficient condition is that a single PR
%iteration fully depolarizes the state, that is, the PR circuit is a
%unitary $1$-design.

\bibitem{Diaconis}
P. Diaconis, Proc. Natl. Acad. Sci. USA {\bf 93}, 1659 (1996).

\bibitem{remark3} The data is noisy due to the two-fold degeneracy of
$\lambda_2$;
%the largest non-trivial eigenvalue of $M'$;
when smoothed, the decay is exponential as expected.

\bibitem{Harrow08}
A. Harrow and R. Low, {\tt quant-ph/0802.1919}.

\end{thebibliography}
\end{document}